\documentstyle[art12,bezier]{article}
\def\b:{\begin{equation}}
\def\b:{\begin{equation}}
\def\b:{\begin{equation}}
\def\e:{\end{equation}}
\def\be:{\begin{eqnarray}}
\def\ee:{\end{eqnarray}}

\def\cl{\centerline}

\def\vi{\vglue 1cm}
\def\vii{\vglue 2cm}

\def\ve{\vfill\eject}

\begin{document}
\baselineskip 19pt
\vii
\cl{\Large{\bf Dual Resonance Model Solves}}
\cl{\Large{\bf the Yang-Baxter Equation}}
\vii
\cl{\large{Satoru SAITO}}
\vi
\cl{\it{Department of Physics, Tokyo Metropolitan University,}}
\cl{\it{Minami-Ohsawa, Hachiohji, Tokyo, Japan\ 192-03}}
\vglue 8cm
\begin{abstract}
The duality of dual resonance models is shown to imply that the four point
string correlation function solves the Yang-Baxter equation. A reduction of
transfer matrices to $A_l$ symmetry is described by a restriction of the
KP $\tau$ function to Toda molecules.
\end{abstract}

\ve
By the old duality I mean the $stu$ duality embodied in the dual resonance
model \cite{DRM} developed in the late 1960's and 1970's as a model which
describes hadron scattering processes. The purpose of this article is to
supply an argument which clarifies the link among three independent subjects
in physics, i.e., string models in particle physics, solvable lattice models
in statistical physics, and soliton theory. In particular I will show in the
following that the old duality assures for a four point string correlation
function to solve the Yang-Baxter equation.

The correspondence between the soliton theory and the string models is rather
straight-forward \cite{SS}. The string correlation functions solve the Hirota
bilinear difference equation (HBDE), defined as \cite{Hirota}
\begin{eqnarray}
&& \alpha f(k_1+1,k_2,k_3)f(k_1,k_2+1,k_3+1)\ \nonumber\\
&& \qquad + \beta f(k_1,k_2+1,k_3)f(k_1+1,k_2,k_3+1)\nonumber\\
&&\qquad\qquad + \gamma f(k_1,k_2,k_3+1)f(k_1+1,k_2+1,k_3)=0
\label{HBDE}
\end{eqnarray}
with $\alpha +\beta+\gamma=0$. This single equation is equivalent to the
KP-hierarchy in soliton theory \cite{Miwa}. A solution of this equation is
called the tau function \cite{DJKM}.

The soliton theory and the solvable lattice models, on the other hand, share
the common structure of integrability, called the quantum inverse scattering
method  \cite{FST}. Very recently a new light was cast into this connection
through the papers \cite{KLWZ}, \cite{Kuniba} in which was pointed out that
the algebraic relation satisfied by the transfer matrix of the solvable lattice
 model with $A_l$ symmetry is nothing but HBDE $(\ref{HBDE})$. Moreover in
\cite{KLWZ} the authors showed that the linear B\"acklund transformation of
HBDE
\cite{SSS} generates a series of Bethe ansatz solutions.

These results can be summarized such that HBDE unifies the three problems in
our consideration. The same solution, however, is interpreted quite
differently from each other, a correlation function of strings, the tau
function of soliton theory, and a transfer matrix of the solvable lattice
models. From mathematical point of view it is apparent that HBDE embodies a
very large symmetry which guarantees integrability of systems with infinite
degrees of freedom irrespective to their physical interpretations. Such a
mathematical unification of different models, however, does not mean
understanding them from the side of physics.

In the case of our concern the link between the string models and the solvable
lattice models is most obscure since their relation is indirect. Apart from
the fact that the correlation functions of strings and the transfer matrix of
solvable models are governed by the same equation, their connection is not
manifest at all. I like to fill this gap by showing that the lattice models,
whose Boltzmann weight is given by the four point correlation function of the
string models, are solvable. The proof is achieved by noticing that the $stu$
duality of the dual resonance models \cite{DRM} guarantees the Yang-Baxter
equation to be solved.

To begin with let us reformulate briefly the string correlation functions in
a way suitable for our discussion in the following \cite{S-Kos,SS}. We
consider $N$ external strings interacting each other through the world sheet
specified by the ground state $|G\rangle$. It is given by
\begin{eqnarray}
&&F_G(K_1,K_2,\cdots,K_N)\nonumber\\
&&:=\langle  0|W(K_1,g_1)W(K_2,g_2)\cdots W(K_N,g_N)|G\rangle.
\label{correlation}
\end{eqnarray}
Here the $j$-th string is supposed to have momentum $K_j^\mu(z)$ distributed
along a path $g_j(z)$ in the world sheet. The path is assumed to close a
contour as the local coordinate $z$ of the world sheet moves around a circle.
The interaction takes place via the vertex operator \cite{DSS}
\begin{eqnarray}
W(K_j,g_j)&=&\exp\left[{1\over 2\pi}\oint {dz\over z}K_j^\mu(z)X_+^\mu(g_j(z))
\right]\nonumber\\
&\times& \exp\left[{1\over 2\pi}\oint {dz\over z}K_j^\mu(z)X_-^\mu(g_j(z))
\right].
\label{vertex W}
\end{eqnarray}
Here the string coordinate $X^\mu(z)=X_+^\mu(z)+X_-^\mu(z)$ is
defined by the following expansion:
\begin{eqnarray}
X_-^\mu(z)&=&x^\mu+\sum_{n=1}^\infty{a_n^\mu\over\sqrt n} z^n,\nonumber\\
X_+^\mu(z)&=&ip^\mu\ln z+\sum_{n=1}^\infty{a_n^{\dagger\mu}\over\sqrt n} z^{-n}
\label{X seibun hyouji}
\end{eqnarray}
whose components satisfy
\begin{equation}
[x^\mu, p^\nu]=i\delta^{\mu\nu},\  [a^\mu_m, {a^\nu_n}^\dagger]=
\delta^{\mu\nu}\delta_{mn},\  m,n\in {\bf Z}_{\ge 1}.
\label{string quantization}
\end{equation}
If $K_j^\mu$ is a constant vector $k_j^\mu$, the vertex operator $W(K_j,g_j)$
turns to the ordinary vertex operator for the external ground state particle
of momentum $k_j^\mu$
\begin{equation}
V(k_j,z_j)=e^{ik_j^\mu X_+^\mu(z_j)}e^{ik_j^\mu X_-^\mu(z_j)}
\end{equation}
where $z_j=g_j(0)$. Therefore $k_j^\mu$ is the barycentric momentum of the
$j$-th string. To make simpler the expressions of formulae the space-time
indices $\mu, \nu, \cdots$ will be suppressed in what follows.

The empty state $|0\rangle$ is defined by
$$
p|0\rangle=a_n|0\rangle=0,\qquad   n=1,2,\cdots,
$$
while the ground state is defined by $|G\rangle=G(X)|0\rangle$, where
\cite{Kato-S}
\begin{eqnarray}
G(X)&=&\theta\Bigl(\zeta-{1\over 2\pi}\oint dX(z)\int^z\omega\Bigr)\nonumber\\
&\times&\exp\Bigl[{1\over 8\pi^2}\oint dX(x)\oint dX(y)\ln{E(x,y)\over x-y}
\Bigr].
\label{G(X)}
\end{eqnarray}
$\theta,\ \omega,\ E(x,y),\ \zeta$ are the Riemann theta function, the Abel
differential, the prime form, and an arbitrary vector, respectively, all
defined on the world sheet of definite number of genus.

The connection of the correlation function $(\ref{correlation})$ with the tau
function of the KP hierarchy was shown \cite{SS} to follow to
\begin{equation}
{F_G(K_1,K_2,\cdots, K_n)\over F_0(K_1,K_2,\cdots, K_n)}=\tau(t).
\label{F_G/F_0=tau}
\end{equation}
Here $F_0$ is given by $(\ref{correlation})$ with $|G\rangle$ replaced by
$|0\rangle$, and $t$ denotes the collection of the soliton coordinates
$\{ t_1,t_2,\cdots\}$ which are related with the string variables by
\begin{equation}
t_n={1\over n}\sum_{j=1}^N{1\over 2\pi}\oint{dz\over z}K_j(z)g_j^n(z),\qquad
n=1,2,\cdots.
\label{t_n=}
\end{equation}
Note that, when $K_j^\mu(z)=k_j^\mu$, this reduces to the Miwa transformation
\cite{Miwa}. The proof that $(\ref{F_G/F_0=tau})$ satisfies $(\ref{HBDE})$ is
exactly the same as in ref.\ \cite{SS}. The variables in $(\ref{HBDE})$ are
any three chosen out of the constant components of $K_j(z)'s$.

We now consider the link between the string model and the solvable lattice
model. The main step toward this problem is to define the Boltzmann weight
properly, so that the Yang-Baxter equation is solved. Here I propose a two
dimensional lattice model whose links are specified by string momenta $K(z)$'s
and Boltzmann weight is given by the four point string correlation function
\begin{eqnarray}
&&R_{K,K'}^{K''',K''}(g,g',\bar g'',\bar g''')\nonumber\\
&&=\langle  0|W(K,g)W(K',g')\bar W(K'',g'')\bar W(K''',g''')|0\rangle.
\label{R}
\end{eqnarray}
Here $\bar W$ is the operator whose in-state and out-state are reversed and
$\bar g(z)=g\left({1\over z}\right)$.
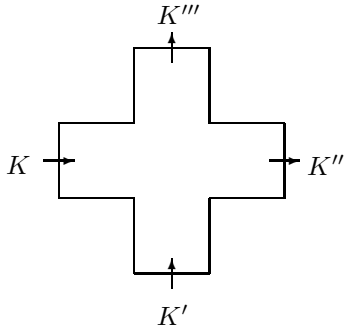
\begin{center}
\begin{minipage}{5cm}\begin{center}\unitlength 1mm\begin{picture}(50,45)
\put(10,20){\line(0,1){10}}
\put(10,20){\line(1,0){10}}
\put(10,30){\line(1,0){10}}
\put(20,30){\line(0,1){10}}
\put(20,40){\line(1,0){10}}
\put(30,30){\line(0,1){10}}
\put(30,30){\line(1,0){10}}
\put(40,20){\line(0,1){10}}
\put(30,20){\line(1,0){10}}
\put(30,10){\line(0,1){10}}
\put(20,10){\line(1,0){10}}
\put(20,10){\line(0,1){10}}
\put(3,23){\makebox(10,3)[l]{$K$}}
\put(23,3){\makebox(10,3)[l]{$K'$}}
\put(23,43){\makebox(10,3)[l]{$K'''$}}
\put(43,23){\makebox(10,3)[l]{$K''$}}
\put(8,25){\vector(1,0){4}}
\put(38,25){\vector(1,0){4}}
\put(25,8){\vector(0,1){4}}
\put(25,38){\vector(0,1){4}}
\end{picture}\end{center}\end{minipage}
\end{center}
Using this Boltzmann weight the transfer matrix of the model is defined by
\begin{equation}
T_{K_1,K_2,\cdots,K_M}^{K'_1,K'_2,\cdots,K'_M}
=\sum_{\{K''_j\}}R_{K''_1,K_1}^{K'_1,K''_2}
R_{K''_2,K_2}^{K'_2,K''_3}\cdots R_{K''_M,K_M}^{K'_M,K''_1}.
\label{T}
\end{equation}
The summation over $K''_j$ means the functional integration over all possible
paths of strings $K''_j$. Hence it turns out to be given explicitly by the
$2M$ point string correlation function, defined on a torus world sheet
associated with the ground state $|G_1\rangle$:
\begin{eqnarray}
=&&\langle  0|W(K_1,g_1)\bar W(K'_1,g'_1)W(K_2,g_2)
\dots\bar W(K'_M,g'_M)|G_1\rangle\nonumber\\
=&&F_{G_1}\left(K_1,K'_1,K_2,K'_2,\cdots,K_M,K'_M\right).
\end{eqnarray}

{}From the transfer matrix $(\ref{T})$ we define an operator ${\cal T}$ simply
by omitting the summation over $K''_1$. According to the standard argument
\cite{Baxter} two of ${\cal T}$ operators with different spectral parameters
commute with each other when the $R$ matrix satisfies the Yang-Baxter equation:
\begin{eqnarray}
&&\sum_{K,K',K''}R_{K_1,K'}^{K_6,K}R_{K',K_2}^{K'',K_3}R_{K,K''}^{K_5,K_4}
\nonumber\\
\qquad &=& \sum_{K,K',K''}R_{K_6,K''}^{K_5,K'} R_{K_1,K_2}^{K'',K}
R_{K,K_3}^{K',K_4}
\label{YB}
\end{eqnarray}
Hence, if $(\ref{YB})$ holds, the lattice model is solvable.
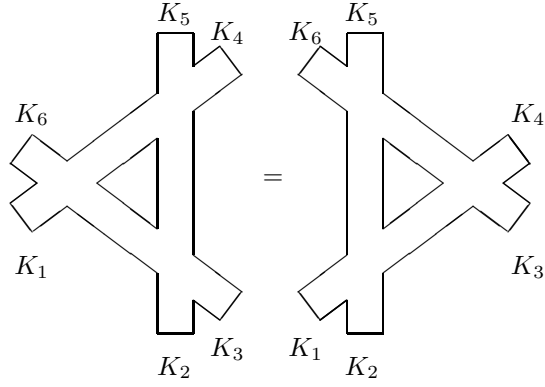
\begin{center}
\begin{minipage}{8cm}\begin{center}\unitlength 1mm\begin{picture}(80,55)
\put(20,10){\line(0,1){8}}
\put(20,18){\line(-4,3){12}}
\put(8,27){\line(-4,-3){4.5}}
\put(0.48,27.36){\line(3,-4){2.88}}
\put(4,30){\line(-4,-3){3.52}}
\put(4,30){\line(-4,3){3.52}}
\put(0.48,32.64){\line(3,4){2.88}}
\put(8,33){\line(-4,3){4.5}}
\put(8,33){\line(4,3){12}}
\put(20,42){\line(0,1){8}}
\put(20,50){\line(1,0){4.8}}
\put(24.8,50){\line(0,-3){4.4}}
\put(24.8,45.6){\line(4,3){3.52}}
\put(28.32,48.24){\line(3,-4){2.88}}
\put(24.8,39.6){\line(4,3){6.5}}
\put(24.8,39.6){\line(0,-1){19}}
\put(24.8,20.4){\line(4,-3){6.5}}
\put(28.32,11.76){\line(3,4){2.88}}
\put(24.8,14.4){\line(4,-3){3.52}}
\put(24.8,14.4){\line(0,-1){4.4}}
\put(12,30){\line(4,3){8}}
\put(12,30){\line(4,-3){8}}
\put(20,24){\line(0,1){12}}
\put(20,10){\line(1,0){4.8}}
\put(34,29){\makebox(3,3)[l]{$=$}}
\put(50,10){\line(0,1){8}}
\put(50,18){\line(4,3){12}}
\put(62,27){\line(4,-3){4.5}}
\put(69.52,27.36){\line(-3,-4){2.88}}
\put(66,30){\line(4,-3){3.52}}
\put(66,30){\line(4,3){3.52}}
\put(69.52,32.64){\line(-3,4){2.88}}
\put(69.52,32.64){\line(-3,4){2.88}}
\put(62,33){\line(4,3){4.5}}
\put(62,33){\line(-4,3){12}}
\put(50,42){\line(0,1){8}}
\put(50,50){\line(-1,0){4.8}}
\put(45.2,50){\line(0,-3){4.4}}
\put(45.2,45.6){\line(-4,3){3.52}}
\put(41.68,48.24){\line(-3,-4){2.88}}
\put(45.2,39.6){\line(-4,3){6.5}}
\put(45.2,39.6){\line(0,-1){19}}
\put(45.2,20.4){\line(-4,-3){6.5}}
\put(41.68,11.76){\line(-3,4){2.88}}
\put(45.2,14.4){\line(-4,-3){3.52}}
\put(45.2,14.4){\line(-4,-3){3.52}}
\put(45.2,14.4){\line(0,-1){4.4}}
\put(58,30){\line(-4,3){8}}
\put(58,30){\line(-4,-3){8}}
\put(50,24){\line(0,1){12}}
\put(50,10){\line(-1,0){4.8}}
\put(1,37.64){\makebox(3,3)[l]{$K_6$}}
\put(1,17.36){\makebox(3,3)[l]{$K_1$}}
\put(20,4){\makebox(3,3)[l]{$K_2$}}
\put(27,6.5){\makebox(3,3)[l]{$K_3$}}
\put(27,48.5){\makebox(3,3)[l]{$K_4$}}
\put(20,51){\makebox(3,3)[l]{$K_5$}}
\put(68.52,37.64){\makebox(3,3)[r]{$K_4$}}
\put(68.52,17.36){\makebox(3,3)[r]{$K_3$}}
\put(46.5,4){\makebox(3,3)[r]{$K_2$}}
\put(39,6.5){\makebox(3,3)[r]{$K_1$}}
\put(39,48.5){\makebox(3,3)[r]{$K_6$}}
\put(46.5,51){\makebox(3,3)[r]{$K_5$}}
\end{picture}\end{center}\end{minipage}
\end{center}
To prove $(\ref{YB})$, I claim that it is nothing but the duality relation.
In fact we first notice that in- and out-states of the string $K(z)$ is
exchanged by the Hermite conjugation of it:
$$
K(z)\quad   \rightarrow\quad    K^\dagger(z)=K\left({1\over z}\right),
$$
from which follows
\begin{equation}
\bar W(K,g)=W(K,\bar g).
\end{equation}
Using this and the fact that a string correlation function is symmetric under
the cyclic permutation the four point function can be rewritten as
\begin{eqnarray}
&&\langle  0|W(K,g)W(K',g')\bar W(K'',g'')\bar W(K''',g''')|0\rangle
\nonumber\\
=&&\langle  0|W(K''',\bar g''')W(K,g)\bar W(K',\bar g')\bar W(K'',g'')|0
\rangle\cr
=&&\langle  0|W(K'',\bar g'')W(K''',\bar g''')\bar W(K,\bar g)\bar W(K',
\bar g')|0\rangle
\end{eqnarray}
Application of these identities yields for the lhs of $(\ref{YB})$
\begin{eqnarray}
&&\sum_{K,K',K''}R_{K_1,K'}^{K_6,K}R_{K',K_2}^{K'',K_3}R_{K,K''}^{K_5,K_4}
\nonumber\\
\qquad &=&\sum_{K,K',K''}R_{K_1,K'}^{K_6,K}R_{K_3,K''}^{K_2,K'}
R_{K_5,K}^{K_4,K''}
\end{eqnarray}
and for the rhs
\begin{eqnarray}
&&\sum_{K,K',K''}R_{K_6,K''}^{K_5,K'}R_{K_1,K_2}^{K'',K}R_{K,K_3}^{K',K_4}
\nonumber\\
\qquad &=&\sum_{K,K',K''}R_{K',K_5}^{K'',K_6}R_{K'',K_1}^{K,K_2}
R_{K,K_3}^{K',K_4}.
\end{eqnarray}
Both of these represent six point string functions defined on the world sheet
of torus. In fact they are identical due to the duality and is given by
\begin{eqnarray}
&&\langle 0|\bar W(K_6,g_6)W(K_5,g_5)\bar W(K_4,g_4)\nonumber\\
&&\qquad\qquad\times W(K_3,g_3)\bar W(K_2,g_2)W(K_1,g_1)|G_1\rangle.
\end{eqnarray}
This justifies the claim.

We have just established the direct link between the string models and the
solvable lattice models. In the rest of this letter I like to demonstrate that
through some reduction we can obtain a familier solvable lattice model. It
will, at the same time, explain the mysterious relation betweem the Yang-Baxter
 equation and HBDE recently discussed in \cite{KLWZ},\cite{Kuniba}.

The transfer matrix $T^{(\mu)}_\nu(\lambda)$ of the solvable lattice model
associated with $A_l$ symmetry was shown \cite{KLWZ},\cite{Kuniba} to solve
HBDE
$(\ref{HBDE})$. The variables $\mu, \nu, \lambda$ of $T^{(\mu)}_\nu(\lambda)$
denote the size of the $\mu\times \nu$ rectangular Young tableux and the
spectral parameter which specifies the Boltzmann weight. They are related
with the variables $k_1, k_2, k_3$ of HBDE according to $\mu=k_2+k_3-1,\
\nu=k_3+k_1-1,\ \lambda=k_1+k_2-1$. This correspondence sounds rather
artificial because $\mu$ and $\nu$ have meaning of size of Young tableaux and
range a certain finite intervals, while $k_1,\ k_2,\ k_3$ range all integers
or periodic boundary conditions are imposed.

In order to resolve this unnaturalness I first write HBDE in terms of the new
variables
\begin{eqnarray}
&& \alpha g(\lambda+1,\mu,\nu)g(\lambda-1,\mu,\nu)\nonumber\\
&&\quad +\ \beta g(\lambda,\mu+1,\nu)g(\lambda,\mu-1,\nu)\nonumber\\
&&\qquad  + \gamma g(\lambda,\mu,\nu+1)g(\lambda,\mu,\nu-1)=0,
\label{HBDE'}
\end{eqnarray}
where $g(\lambda,\mu,\nu)=f(k_1,k_2,k_3)$. In the following I remark the
results quoted from our recent work \cite{toda30} in a slightly different form,
 appropriate in our discussion.

\noindent
{\it Remark :

Let $g(\lambda,\mu,\nu)$ be a solution of HBDE $(\ref{HBDE'})$, and
$\mbox{\boldmath$A$}(\bar\lambda,\bar\mu,\bar\nu)$ an octahedron consisting of
 the nearest neighbours of the point at $(\lambda,\mu,\nu)=(\bar\lambda,
\bar\mu,\bar\nu)$ in the lattice space ${\mbox{\boldmath$Z$}}^3$, then
\begin{equation}
\bar g(\lambda,\mu,\nu)=\cases{g(\lambda,\mu,\nu) ,\ (\lambda,\mu,\nu) \in
{\mbox{\boldmath$A$}}(\bar\lambda,\bar\mu,\bar\nu)\cr
 0\quad {\rm otherwise}\cr}
\label{Toda atom}
\end{equation}
is also a solution to HBDE $(\ref{HBDE'})$.}

\noindent
This is the smallest piece of Toda lattice which is shown in Fig. a. The proof
of $(\ref{Toda atom})$ is simple. Namely consider another octahedron
$\mbox{\boldmath$A$}'$ which shares at least one point of \mbox{\boldmath$A$}.
Since $g(\lambda,\mu,\nu)=0$ on every lattice point surrounding
\mbox{\boldmath$A$}, $g(\lambda,\mu,\nu)'s$ on the octahedron
$\mbox{\boldmath$A$}'$ automatically satisfy $(\ref{HBDE'})$.

\begin{center}\begin{minipage}{5cm}\unitlength 1mm\begin{picture}(50,50)
\thicklines
\put(30,15){\circle*{2}}\put(25,25){\circle*{2}}\put(15,30){\circle*{2}}
\put(45,30){\circle*{2}}\put(35,35){\circle*{2}}\put(30,45){\circle*{2}}
\put(30,30){\circle{2}}
\put(15,30){\line(1,1){15}}\put(15,30){\line(1,-1){15}}
\put(15,30){\line(2,-1){10}}\put(25,25){\line(1,4){5}}
\put(25,25){\line(1,-2){5}}\put(25,25){\line(4,1){20}}
\put(45,30){\line(-1,1){15}}\put(45,30){\line(-1,-1){15}}
\bezier{15}(35,35)(33,39)(30,45)\bezier{15}(35,35)(39,33)(45,30)
\bezier{30}(35,35)(31,34)(15,30)\bezier{30}(35,35)(34,31)(30,15)
\put(10,15){\line(0,1){5}}\put(10,15){\line(1,0){5}}
\put(10,15){\line(-1,-1){4}}\put(9,21){\makebox(2,2)[l]{$\lambda$}}
\put(3,8){\makebox(2,2)[l]{$\mu$}}
\put(16,14){\makebox(2,2)[l]{$\nu$}}
\put(20,0){\makebox(10,2)[l]{Fig. a}}
\end{picture}\end{minipage}\end{center}

The generalization of $(\ref{Toda atom})$ to an arbitrary size of piece of
Toda lattice is straightforward. Let us call such a piece a Toda molecule
according to ref.\cite{Toda molecule}. Then the smallest unit
$(\ref{Toda atom})$ should be called a Toda atom. A Toda molecule must be
rectangular when it is sliced perpendicular to each axis of the lattice, for
it being a solution of HBDE $(\ref{HBDE'})$. We can consider a collection of
Toda molecules if they are disjoined with each other. An example of a slice of
such a collection is given in Fig. b. Note that each piece can be an
independent solution of HBDE.
\begin{center}\begin{minipage}{8cm}\unitlength 1mm\begin{picture}(80,55)
\thicklines
\put(10,5){\circle{1}}\put(15,5){\circle{1}}\put(20,5){\circle{1}}
\put(25,5){\circle{1}}\put(30,5){\circle{1}}\put(35,5){\circle{1}}
\put(40,5){\circle{1}}\put(45,5){\circle{1}}\put(50,5){\circle{1}}
\put(55,5){\circle{1}}\put(60,5){\circle{1}}\put(65,5){\circle*{1}}
\put(70,5){\circle{1}}
\put(10,10){\circle{1}}\put(15,10){\circle{1}}\put(20,10){\circle{1}}
\put(25,10){\circle*{1}}\put(30,10){\circle{1}}\put(35,10){\circle{1}}
\put(40,10){\circle*{1}}\put(45,10){\circle{1}}\put(50,10){\circle{1}}
\put(55,10){\circle{1}}\put(60,10){\circle*{1}}\put(65,10){\circle{1}}
\put(70,10){\circle*{1}}
\put(10,15){\circle{1}}\put(15,15){\circle{1}}\put(20,15){\circle*{1}}
\put(25,15){\circle{1}}\put(30,15){\circle*{1}}\put(35,15){\circle*{1}}
\put(40,15){\circle{1}}\put(45,15){\circle*{1}}\put(50,15){\circle{1}}
\put(55,15){\circle{1}}\put(60,15){\circle{1}}\put(65,15){\circle*{1}}
\put(70,15){\circle{1}}
\put(10,20){\circle{1}}\put(15,20){\circle*{1}}\put(20,20){\circle{1}}
\put(25,20){\circle*{1}}\put(30,20){\circle{1}}\put(35,20){\circle{1}}
\put(40,20){\circle*{1}}\put(45,20){\circle{1}}\put(50,20){\circle{1}}
\put(55,20){\circle{1}}\put(60,20){\circle{1}}\put(65,20){\circle{1}}
\put(70,20){\circle*{1}}
\put(10,25){\circle{1}}\put(15,25){\circle{1}}\put(20,25){\circle*{1}}
\put(25,25){\circle*{1}}\put(30,25){\circle{1}}\put(35,25){\circle{1}}
\put(40,25){\circle{1}}\put(45,25){\circle{1}}\put(50,25){\circle{1}}
\put(55,25){\circle*{1}}\put(60,25){\circle*{1}}\put(65,25){\circle{1}}
\put(70,25){\circle{1}}
\put(10,30){\circle{1}}\put(15,30){\circle{1}}\put(20,30){\circle*{1}}
\put(25,30){\circle{1}}\put(30,30){\circle*{1}}\put(35,30){\circle{1}}
\put(40,30){\circle{1}}\put(45,30){\circle{1}}\put(50,30){\circle*{1}}
\put(55,30){\circle*{1}}\put(60,30){\circle*{1}}\put(65,30){\circle*{1}}
\put(70,30){\circle{1}}
\put(10,35){\circle{1}}\put(15,35){\circle*{1}}\put(20,35){\circle{1}}
\put(25,35){\circle*{1}}\put(30,35){\circle{1}}\put(35,35){\circle*{1}}
\put(40,35){\circle{1}}\put(45,35){\circle{1}}\put(50,35){\circle*{1}}
\put(55,35){\circle*{1}}\put(60,35){\circle*{1}}\put(65,35){\circle{1}}
\put(70,35){\circle*{1}}
\put(10,40){\circle*{1}}\put(15,40){\circle{1}}\put(20,40){\circle*{1}}
\put(25,40){\circle{1}}\put(30,40){\circle*{1}}\put(35,40){\circle{1}}
\put(40,40){\circle{1}}\put(45,40){\circle{1}}\put(50,40){\circle{1}}
\put(55,40){\circle*{1}}\put(60,40){\circle{1}}\put(65,40){\circle*{1}}
\put(70,40){\circle{1}}
\put(10,45){\circle{1}}\put(15,45){\circle*{1}}\put(20,45){\circle{1}}
\put(25,45){\circle*{1}}\put(30,45){\circle{1}}\put(35,45){\circle{1}}
\put(40,45){\circle{1}}\put(45,45){\circle{1}}\put(50,45){\circle{1}}
\put(55,45){\circle{1}}\put(60,45){\circle*{1}}\put(65,45){\circle{1}}
\put(70,45){\circle{1}}
\put(10,50){\circle{1}}\put(15,50){\circle{1}}\put(20,50){\circle*{1}}
\put(25,50){\circle{1}}\put(30,50){\circle{1}}\put(35,50){\circle{1}}
\put(40,50){\circle{1}}\put(45,50){\circle{1}}\put(50,50){\circle{1}}
\put(55,50){\circle{1}}\put(60,50){\circle{1}}\put(65,50){\circle{1}}
\put(70,50){\circle{1}}
\put(25,10){\line(1,1){5}}\put(25,10){\line(-1,1){10}}
\put(30,15){\line(-1,1){10}}\put(20,15){\line(1,1){5}}
\put(15,20){\line(1,1){5}}
\put(40,10){\line(1,1){5}}\put(40,10){\line(-1,1){5}}
\put(40,20){\line(1,-1){5}}\put(40,20){\line(-1,-1){5}}
\put(60,10){\line(1,-1){8}}\put(60,10){\line(1,1){10}}
\put(65,15){\line(1,-1){8}}\put(70,20){\line(1,-1){3}}
\put(65,5){\line(1,1){8}}\put(70,20){\line(1,-1){3}}
\put(25,25){\line(1,1){10}}\put(25,25){\line(-1,1){15}}
\put(30,30){\line(-1,1){15}}\put(35,35){\line(-1,1){15}}
\put(20,30){\line(1,1){10}}\put(15,35){\line(1,1){10}}
\put(10,40){\line(1,1){10}}
\put(60,25){\line(1,1){10}}\put(60,25){\line(-1,1){10}}
\put(65,30){\line(-1,1){10}}\put(70,35){\line(-1,1){10}}
\put(55,30){\line(1,1){10}}\put(50,35){\line(1,1){10}}
\put(55,25){\line(-1,1){5}}\put(55,25){\line(1,1){5}}
\put(55,35){\line(1,-1){5}}\put(55,35){\line(-1,-1){5}}
\put(10,0){\makebox(10,0)[l]{Fig. b\quad A slice of Toda molecules}}
\end{picture}\end{minipage}\end{center}

Now we go back to our lattice model whose Boltzmann weight is given by
$R^{K''',K''}_{K,K'}$ of $(\ref{R})$ but $K'$ and $K'''$ are reduced to their
 barycentric momenta taking only integral numbers. Using this Boltzmann weight
we construct the transfer matrix which carries only integral numbers in its
legs. According to the above remarks we can consider any size of Toda molecules
 in equal basis. From this point of view we are allowed to think of these
numbering of the legs as specifying the size of the rectangular slice of a
molecule instead of address on the lattice space.

Adopting this convention we consider a solvable lattice model with the $A_l$
symmetry. We identify the transfer matrix associated with the $\mu\times
\nu=(k_2+k_3-1)\times (k_3+k_1-1)$ rectangular type Young tableux with the
same size of Toda molecule in the $(\mu,\nu)$ plane. Further identification of
 the spectral parameter $\lambda$ with $k_1+k_2-1$ completes the desired
correspondence.

Before closing this note I like to mention a few comments. The partition
function of our
lattice model itself is a correlation function $(\ref{correlation})$ of
strings. Hence it is a solution of HBDE. This form of general string
correlation function was calculated \cite{WWET} explicitly to reproduce
$(\ref{correlation})$ using the vertex operator $W$ in $(\ref{vertex W})$ as
a building block. From this point of view the solvable lattice model is
nothing but a special case of analogue (or fish net) models \cite{Sakita}
discussed in connection with hadron scattering processes. Therefore this
model has been shown integrable in two folds. Namely it satisfies
 HBDE \cite{SS} and also satisfies the Yang-Baxter equation as it was shown
in this paper.

The braid of strings was discussed in \cite{Cateau-S}. There the vertex
operator $W$ was regarded as representing a state of string, and a braid of
strings was caused through an exchange of the order of $W$'s. The exchange
matrix was derived assuming that states of the strings were not changed under
their exchange of oder because of the duality. Hence it is included as a
special case of present work.

{\small
This work is supported in part by the Grant-in-Aid for general Scientific
Research from the Ministry of Education, Science, Sports and Culture, Japan
(No.06835023), and the Fiscal Year 1996 Fund for Special Research Projects
at Tokyo Metropolitan University.}

\end{document}